\documentclass[twocolumn,prb,showpacs]{revtex4}
\usepackage{graphicx}

\def\ha{Hamiltonian }
\def\<{\langle}
\def\>{\rangle}
\def\bpsi{\bar{\psi}}
\def\sx{\sigma_x}

\def\sz{\sigma_z}
\def\twid{\tilde{\Delta}}
\def\ptau{\tau^\prime}
\def\half{{\frac{1}{2}}}
\def\be{\begin{equation}}
\def\ee{\end{equation}}

\begin{document}
\preprint{cond-mat} \title{Entanglement of a qubit with a single
oscillator mode}

\author{Gregory Levine}

\address{Department of Physics and Astronomy, Hofstra University,
Hempstead, NY 11549}

\author{V. N. Muthukumar}
\address{Department of Physics, Princeton University, 
Princeton, NJ 08544}
\date{\today}

\begin{abstract}
We solve a model of a qubit strongly coupled to a massive
environmental oscillator mode where the qubit backaction is treated
exactly.  Using a Ginzburg-Landau formalism, we derive an effective
action for this well known localization transition. An
entangled state emerges as an instanton in the collective
qubit-environment degree of freedom and the resulting model is shown to
be formally equivalent to a Fluctuating Gap Model (FGM) of a
disordered Peierls chain.  Below the transition, spectral weight is
transferred to an exponentially small energy scale leaving the qubit
coherent but damped. Unlike the spin-boson model, coherent and effectively localized behaviors may coexist.

\end{abstract}

\maketitle

The subject of quantum computing has led to renewed interest in the
theory of quantum decoherence and also exposed a practical side to the
theory.  Scalable, persistent current designs for qubits
involve measurement devices that are permanently coupled, leading to a
continuous dephasing of the qubit \cite{vdw,devoret}.  Before a
quantum measurement can be made, the qubit must become entangled with
the measurement device.  The conflicting demands of quantum
measurement and minimal dephasing is an active topic of study within
the theory of mesoscopic quantum detectors
\cite{schon_rmp,averin_wcm,girvin_02,mozyrsky}.

In this letter, we study the problem of how a qubit becomes entangled
with a single environmental mode when the qubit backaction on the
environment is not neglected.  This is perhaps the simplest quantum
environment, but already displays great complexity.  Models of
dissipative quantum environments such as the spin-boson model (SBM)
are intentionally weakly coupled and therefore neglect backaction.
However, environments imposed by quantum detectors may be strongly
coupled and exhibit sharp spectral features. The entanglement described here might be realized in any environment (e.g detector or qubit-qubit coupler) that
exhibits a low energy monochromatic spectrum; as an example, we
discuss the environment imposed by an underdamped DC-SQUID detector.
Lastly, we point out the equivalence between the present model and a
particular model of disordered fermions.

 Consider a qubit coupled to a single harmonic oscillator described by
the \ha
\begin{equation}
\label{ham}
H = \Delta \sx + \lambda \frac{1}{\sqrt{2m\omega}} (a + a^\dagger) \sz
+ \omega a^\dagger a
\end{equation}
If the oscillator is replaced by a classical, adiabatic oscillator
(the limit $\omega \rightarrow 0$, $m\omega^2 \rightarrow
\textrm{const}$), it is well known \cite{weiss_book} that (\ref{ham})
exhibits a bifurcation at $\lambda^2 = \Delta m \omega^2$ in that the
energy of the ground state is now minimized by assuming a nonzero
value of the oscillator displacement, $x=\pm x_0$.  In the doubly
degenerate ground state, the qubit is localized with a nonzero
expectation value, $\<\sz\>\ne 0$.  The classical calculation begs two
questions: 1) If the oscillator localizes the qubit, is all coherent
behavior destroyed?  2) If the qubit and oscillator {\sl weakly}
interact ($\lambda^2 << \Delta m \omega^2$) aren't they always
entangled to some degree?

We show that the onset of entanglement with the oscillator
displacement becomes {\sl discontinuous} in the massive limit, with a
component of the ground state wavefunction playing the role of an
order parameter in a second order phase transition. Below the
transition, damped coherent behavior and entangled, effectively
localized behavior coexist.

Entanglement of the qubit in our case is distinct from that of
weak-coupling theories (such as the SBM) in that the ergodicity of the
environment is effectively broken and fluctuations become
non-Gaussian.  This latter effect, of course, cannot be seen by
integrating out the environmental degrees of freedom to find the
effective action of the qubit.  In this respect, our calculation bears
some relation to the description of quantum measurement by Single
Electron Transistor (SET) and overdamped DC-SQUID detectors
\cite{schon_rmp,averin_wcm}.  In this scheme, the qubit degrees of
freedom are integrated out to yield the response function of the
detector. 

To represent the qubit, we choose the finite temperature formalism of
Popov and Fedotov \cite{popov_88} in which the action is written in terms of
spinors satisfying modified boundary conditions.  The imaginary time
action corresponding to (\ref{ham}) is
\begin{equation}
\label{action}
S = \int{d\tau \bpsi(\tau) (\partial_\tau + \Delta \sx + \lambda
x(\tau)\sz)\psi(\tau)} + S_0
\end{equation}
where $S_0 = \int{d\tau\left(\frac{1}{2}m\dot{x}^2(\tau) + \half k
x^2(\tau) \right)}$, $k\equiv m \omega^2$ and $\psi(\tau)$ is a spinor
\begin{equation} 
\psi(\tau) = 
\left(\begin{array}{c}
      \psi_\uparrow(\tau)  \\  \psi_\downarrow(\tau)     
\end{array} \right)
\end{equation}

The spin-1/2 is formally represented as a fermion spinor with an
imaginary chemical potential. The latter eliminates unphysical
fermionic states from the Hilbert space.  A gauge transformation
removes the chemical potential and shifts the Matsubara frequencies to
$\omega_n = 2\pi(n+1/4)/\beta$. Conventional finite temperature field
theory techniques may now be applied to the action (\ref{action})
\cite{popov_88}. The fermions are now integrated out to yield an
effective boson action:
\begin{eqnarray}
Z & = &\int{D\bpsi D\psi Dx e^{-S}} \propto
\int{Dx e^{-S_{\rm{eff}}}} \\  S_{\rm{eff}} & = & -\rm{tr}
\log{(-G_0^{-1} + \frac{\lambda}{\sqrt{\beta}}x(\omega_n -
\omega_m)\sz)} + S_0
\end{eqnarray}
where $G_0 = (i\omega_n - \Delta\sx)^{-1}$ is the free fermion Green's
function. The coordinate $x(\tau)$ now represents a collective
coordinate for $\sz$ and the oscillator displacement.

Expanding the action to quartic order in $\lambda$ and regrouping into
dynamic and static terms, respectively, we obtain:
\begin{eqnarray}
\nonumber
S_{\rm eff} &=&  \half \sum{(m p_n^2 + \pi(p_n) - \pi(0))|x(p_n)|^2} \\
&+& \half  \sum{(k + \pi(0))|x(p_n)|^2}\\
\nonumber
&+& \sum{D(0)|x(p_n)|^4}
\end{eqnarray}
where the coefficients are given by
\begin{eqnarray}
\nonumber \pi(p_n) & = &
\frac{\lambda^2}{\beta}\textrm{tr}\sum{G_0(p_n + \omega_m)\sz
G_0(\omega_m)\sz} \\  & = & -\lambda^2 \tanh{\beta
\Delta}\frac{4\Delta}{p_n^2+4\Delta^2}\\ \nonumber D(0) & =
&\frac{\lambda^4}{\beta^2} \textrm{tr}\sum{\lbrack
G_0(\omega_m)\sz\rbrack^4} = \frac{\lambda^4}{8\beta
\Delta^3}\tanh{\beta \Delta}
\end{eqnarray}
Owing to the modified Matsubara frequencies, these sums are computed
with a modified Fermi distribution, $f(\omega) = (ie^{\beta \omega} +
1)^{-1}$. Defining the dimensionless control parameter $\alpha =
\frac{\lambda^2}{k \Delta}$, the effective action is transformed back
to imaginary time at zero temperature.
\begin{equation}
S_{\rm eff}
=\int{d\tau\left(\half(m+\frac{\lambda^2}{4\Delta^3})\dot{x}^2+\half k
(1-\alpha)x^2 + \frac{\lambda^4}{8\Delta^3} x^4 \right)}
\end{equation}
The mass is enhanced by a dynamical quantity and, most significantly,
the action exhibits an instability at $\alpha=1$.  When $\alpha >1$
and $\omega = 0$, the action is minimized for $x_0 = \pm
\frac{\Delta}{\lambda}(\frac{2\alpha -2}{\alpha})^{1/2}$.  For $\omega
\neq 0$, there is no true broken symmetry, however an approximate
calculation of the instanton action
\begin{equation}
\label{S_I}
S_{\rm i} = \sqrt{2}
\frac{\Delta}{\omega}\frac{(\alpha-1)^{3/2}}{\alpha^2} \sqrt{1 +
\frac{\omega^2}{\Delta^2}\alpha}
\end{equation}
shows that, although the symmetry changes at $\alpha >1$, the
instanton is only stabilized for $\Delta/\omega >> (\alpha
-1)^{-3/2}$.  This defines the entangled regime of the
qubit and oscillator in that the ground state is a coherent
superposition of two states of the collective coordinate, $x(\tau) =
\pm x_0$.

Now we must find the corresponding dynamics of the qubit, in
particular, the spin-spin correlation function.  It may be shown that
\begin{equation}
\< T_\tau \sigma_i(\tau) \sigma_j(0) \> = \frac{1}{Z} \int{Dx\< T_\tau
\bpsi \sigma_i \psi \bpsi \sigma_j \psi
\>_{S_{\rm{f}}[x]}e^{-S_{\rm{eff}}}}
\end{equation}
where $S_{\rm{f}}[x]$ denotes averaging over the fermionic action for
a particular realization of the boson field $x(\tau)$.  By Wick
contractions, this average is related to the single fermion Green's
function, in the presence of an inhomogeneous field $x(\tau)$, which
solves the equation of motion: $(\partial_\tau + \Delta \sx + \lambda
x(\tau)\sz)G(\tau \tau^\prime;x(\tau)) = \delta(\tau \tau^\prime)$.
This approach has been used to implement bosonization in 1-d fermion
systems \cite{lee_88}. In the present case, the Green's functions must
satisfy the boundary conditions implied by the shifted Matsubara
frequencies: $G(\tau + \beta) = -iG(\tau)$.  Furthermore, the boson
field $x$ must be written as the sum of an instanton trajectory and a
small oscillation: $x(\tau) = x_{\rm i}(\tau) + r(\tau)$. We now make
the non-Abelian Schwinger ansatz \cite{kopietz_99},
\begin{equation}
\label{Gansatz}
G(\tau \ptau;x(\tau)) = U(\tau)g(\tau \ptau;x_{\rm
i}(\tau))U^{-1}(\ptau),
\end{equation}
where $g$ satisfies the equation of motion but with the field
$x(\tau)$ restricted to the instanton trajectory: $(\partial_\tau +
\Delta \sx + \lambda x_{\rm i}(\tau)\sz)g(\tau \tau^\prime;x_{\rm
i}(\tau)) = \delta(\tau \tau^\prime)$, and $x_{\rm i}(\tau) = \pm
x_0$.  Now $U(\tau)$ must satisfy the auxillary condition
\begin{equation}
\label{auxcond}
\partial_\tau U + [\tilde{H},U] = -\lambda r(\tau) \sz U
\end{equation}
where $\tilde{H} \equiv \Delta\sx + \lambda x_{\rm i}(\tau) \sz$.

A solution of the equation of motion for $g$ which satisfies the
boundary conditions is obtained for all $\tau$ away from the instanton
kinks (consistent with the dilute gas approximation); for $\tau<\ptau$
\begin{eqnarray}
\nonumber g(\tau\ptau;x_{\rm i}(\tau)) & = & \half f(\twid)\left[1 +
\frac{\Delta}{\twid}\sx + \frac{\lambda
x_{\rm i}(\tau)}{\twid}\sz\right]e^{-\twid(\tau-\ptau)} \\ \nonumber & + &
\half f(-\twid)\left[1 - \frac{\Delta}{\twid}\sx - \frac{\lambda
x_{\rm i}(\tau)}{\twid}\sz\right]e^{\twid (\tau-\ptau)}
\end{eqnarray}
where $\twid$ is the Rabi energy $\twid = \sqrt{\Delta^2 + \lambda^2
x_0^2}$.  The Green's function $g(\tau>\ptau)$ is obtained by the
replacement $f(z) \rightarrow -\bar{f}(z)$ where $\bar{f}(z) \equiv
f^*(-z)$.  The auxilliary condition (\ref{auxcond}) leads to a set of
Riccati equations for an appropriate parameterization of $U$ which are
difficult to solve generally \cite{kopietz_99}.  In the strong
coupling limit (i.e. to lowest order in $\Delta/\twid$) the
interacting Green's function is found to be $G(\tau \ptau; x(\tau)) =
g(\tau \ptau; x_{\rm i}(\tau)) \exp{(-\lambda \sz
\int_{\tau}^{\ptau}{r(\tau)d\tau})}$.

Within this scheme we need to evaluate $c_{zz}(\tau-\tau^\prime) = \<
T_\tau \sigma_z(\tau) \sigma_z(\tau^\prime) \>_{S_{\rm{f}}[x]}$; other
correlation functions are obtained similarly.
$c_{zz}(\tau-\tau^\prime)$ involves two contractions yielding a DC
part of $c_{zz}$ and a part at energy $2 \twid$, respectively.  These
correlation functions must in turn be averaged over the effective
boson action, $S_{\rm eff}$.
\begin{eqnarray}
\label{int_g}
\nonumber\< T_\tau \sigma_z(\tau) \sigma_z(\ptau) \> & =
&\<\lim_{\tau_1 \to \tau^+} \rm{tr}\sz G(\tau_1 \tau;x(\tau_1))\\ &
\times & \lim_{\tau_2 \to \tau^{\prime +}} \rm{tr}\sz G(\tau_2
\ptau;x(\tau_2))\>_{S_{\rm eff}}\\ \nonumber & + & \< \rm{tr} \sz
G(\tau \ptau;x(\tau)) \sz G(\ptau \tau;x(\ptau)) \>_{S_{\rm eff}}
\end{eqnarray}

The first contraction in (\ref{int_g}) will involve products of
$x_{\rm i}(\tau)$ at different times, giving the proper long time
behavior; the second contraction will involve the average over boson
fluctuations $r(\tau)$ that dress the qubit oscillations at energy
$\twid$. To evaluate the average over $S_{\rm eff}$ we need both boson
correlation function.  Within the instanton approximation, the first
contraction in (\ref{int_g}) involves
\begin{equation}
\label{inst_corr}
\<T_\tau x_{\rm i}(\tau)x_{\rm i}(\ptau)\>_{S_{\rm
eff}} = x_0^2 e^{-\Gamma|\tau-\ptau|} 
\end{equation}
where $\Gamma = \omega\sqrt{S_{\rm i}/2\pi}\exp{-S_{\rm i}}$.  The
second contraction in (\ref{int_g}) involves the average over
fluctuations of $r(\tau)$, $\<\cosh{\int{d\tau r(\tau)}}\>_{S_{\rm
eff}}$.  Averages of this form and their fourier transforms have been
discussed in several places \cite{mahan}.  The
relationship to the FGM \cite{fgm} may now be seen: Replacing (in
ref. \cite{kopietz_99}) the 1-d spatial degree of freedom in the FGM
by imaginary time, eqn. (\ref{action}) is the action for a fermion of
frequency $i\Delta$ propagating in a Peierls chain with a gap function
$\lambda x(\tau)$. Averaging (in eqn. (\ref{int_g})) over the
instanton fluctuations of $x(\tau)$ with correlator (\ref{inst_corr})
is equivalent to the disorder average in ref. \cite{kopietz_99} with a
correlation length given by the instanton time, $\Gamma^{-1}$.  In the
FGM, the density of states at the Fermi surface is accessed by the
zero frequency limit $\omega \rightarrow i0^+$ which corresponds to
$\Delta \rightarrow 0^+$, the strong coupling limit in the present
work.  The two models differ in that the correlation length of the gap
disorder in the FGM is an independently controlled parameter whereas
the instanton fluctuations are generated spontaneously in the present
model.  In addition, the fermions in the present model satisfy twisted
boundary conditions.

Continuing to real frequencies and denoting the imaginary part of the
fourier transform of (\ref{int_g}) by
$s_{zz}(\nu)$, we obtain for $\alpha >1$ a correlation function of the
form
\begin{eqnarray}
\label{spectral}
\nonumber s_{zz}(\nu) &=& (\frac{\lambda x_0}{\twid})^2
\delta(\nu - \Gamma)\\ &+& \frac{1}{2}(1 - (\frac{\lambda
x_0}{\twid})^2 + (\frac{\Delta}{\twid})^2)\\ \nonumber &\times&
\sum_{mn}{\frac{p^{n+m} e^{-2p}}{m!n!}\delta(\nu - 2\twid
-(n-m)\omega)})
\end{eqnarray}
for $\nu>0$ where $p \equiv \alpha \frac{\Delta}{\omega}$. This is our
main result. Equation (\ref{spectral}) shows that for the broken
symmetry phase ($\alpha >1$) two distinct spectral features are
formed. The low energy feature with weight $w_{\rm ent} =
\frac{1}{2}(\lambda x_0/\twid)^2 \simeq \frac{\alpha-1}{\alpha}$ (at
energy $\Gamma$) corresponds to the qubit entangled with the
oscillator and in a superposition of the form $|+ ; +x_0\> \pm |- ;
-x_0\>$, where $|+/-\>$ refer to states of the qubit with nonvanishing
$z$-polarization.  The high energy set of delta functions (at energy
$\sim 2\twid$), corresponds to the decoupled (unentangled) but
dephased qubit with weight $w_{\rm free} \simeq 1/\alpha$.  Although
the spectrum is discrete, dephasing may be estimated by considering
the envelope of the sidebands associated with the primary
resonance. For small $\omega$ (large $p$) the Poisson distributions
become approximately Gaussian and the dephasing rate is estimated to
be $\omega p^{1/2}$; the fractional width of the resonance, $1/Q$, is
then $(\alpha \frac{\omega}{\Delta})^{1/2}$.  The qubit becomes
critically damped when this factor is unity.  When $m \rightarrow
\infty$, keeping $k$ and therefore $\alpha$ constant, the width of the
resonance goes to zero as $O(m^{-1/4})$ (although the number of bosons
diverges), in agreement with the classical calculation.

The spectral weights appearing in equation (\ref{spectral}) satisfy
the sum rule: $w_{\rm ent} + w_{\rm free} = 1$.  At critical coupling,
a redistribution of spectral weight is initiated and spectral weight
flows from the coherent feature to the entangled one.  Unlike the
SBM where the system spin becomes overdamped before it
becomes localized, coherent and effectively
localized behavior coexist in the present model\cite{levine_01}.
\begin{figure}
\includegraphics[width=6cm]{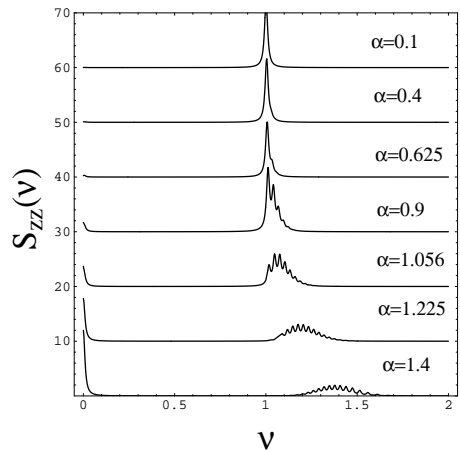}
\caption{\label{fig1} Numerical diagonalization of Hamiltonian
(\ref{ham}). Dynamical spin-spin correlation function showing the
emergence of the low energy instanton feature corresponding to
entanglement.  In this calculation, $\omega=0.01, \Delta=0.5,
k=0.2$. Peaks are artificially broadened and vertical scale on curves
for the largest four values of $\alpha$ is magnified $\times 2$.}
\end{figure}
An approximate numerical diagonalization of the \ha (\ref{ham})
confirm these results.  Figure 1 shows that a feature of energy
$~\Gamma$ appears in the correlation function $s_{zz}(\nu)$ for
$\alpha \sim 1$, while the primary resonance remains distinct from the
instanton feature. As expected from the effective action $S_{\rm
eff}$, $w_{\rm ent}^{1/2}$ is the order parameter for a second order
phase transition; in the strict adiabatic limit ($\Delta/\omega
\rightarrow \infty$), the derivative $d w_{\rm ent}^{1/2}/d\alpha$
would diverge.  To illustrate the sharp onset of entanglement, we
have computed the entanglement entropy $S_{\rm e} \equiv -{\rm tr}\rho
\log_2{\rho}$, where $\rho$ is the ground state reduced density matrix
of the qubit.  The entropy, shown in fig. 2, is seen to sharply
increase from zero at the onset of entanglement
($\alpha=1$). Similarly to the order parameter, the slope of this
graph appears to have a discontinuity at $\alpha=1$ as $\Delta/\omega \rightarrow
\infty$. This is not unexpected, noting that thermodynamic
entropy is given in terms of the Gibbs free energy by $S_{\rm th}=
-\partial G/\partial T$.  Taking $G$ to be the static limit of $S_{\rm
eff}$, $-\partial G/\partial \alpha= \frac{1}{2}k x_0^2$, which is
proportional to the square of the order parameter. Thus,
entanglement entropy appears to behave analogously to thermodynamic
entropy in this second order phase transition.

As a potential example of this transition, we now turn to experimental
realizations of superconducting qubits.  In \cite{vdw} it was
demonstrated that the dominant source of decoherence in a persistent
current (phase) qubit was the electromagnetic environment of the
DC-SQUID, characterized \cite{leggett_leshouches} by an RL$_{\rm J}$C
impedance $Z(\omega)$ (L$_{\rm J}$ is the Josephson equivalent
inductance.) Dephasing follows from the spin-boson model where
$J(\omega) \propto \frac{1}{\omega}{\rm Re}Z(\omega)$; in particular,
$J$ is ohmic at low frequencies.
\begin{figure}
\includegraphics[height=4.2cm,width=7cm]{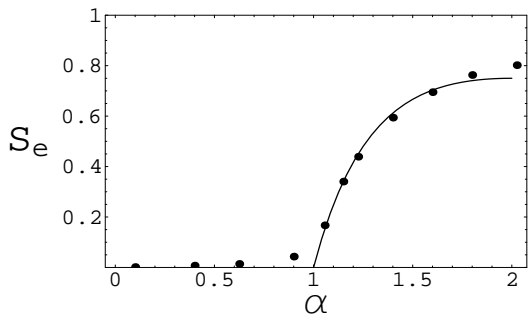}
\caption{\label{fig2} Entanglement entropy computed numerically from
Hamiltonian (\ref{ham}) with $\omega=0.002, \Delta=0.5, k=0.2$. Solid line is a plot of $x_0^2$ scaled by a constant factor.}
\end{figure}
However, $J(\omega)$ is highly structured, dominated by
the plasma mode at the frequency $\omega = 1/\sqrt{L_{\rm J}C}$ with a
$Q \sim 10$ for typical parameters \cite{vdw}.  The $\sz$ coupling
to the qubit comes from the circulating component of current in the
DC-SQUID.  The effective coupling
constant, $\lambda$, reflecting the mutual inductance ($M$) between
the persistent current in the qubit ($I_p \sz$) and circulating
current of the DC-SQUID is then $\lambda = 2MI_0I_p \sin{\half
\bar{\phi}}\sin{\bar{\theta}}$, where $I_0$ is the critical current, $\bar{\theta}$ is the net Josephson phase and $\bar{\phi} \equiv 2\pi (\Phi/\Phi_0)$ the dimensionless flux at the operating point.   Linearizing the effective potential of the SQUID about the operating point, the stiffness is found to be  $k = \frac{\Phi_0}{2
\pi}\sqrt{4 I_0^2 \cos^2{\half \phi} - I_e^2}$, where $I_e$ is the external current bias. Thus the effective potential softens and the dimensionless coupling constant
$\alpha=\frac{\lambda^2}{k \Delta}$ increases as $I_e$ approaches the
switching value.  

To illustrate our point, we use $M=20$ pH, $I_p=900$ nA, $I_0=150$
nA, and fix $I_e$ at $0.95$ of
the critical switching value, $2I_0\cos{\half \bar{\phi}}$, to obtain
a dimensionless coupling of $\alpha \sim 1$.  Once entangled, two
distinct coherent oscillations in $\sz$ would be observed: the primary oscillation at
$2\Delta$ and the entangled one at $\Gamma$.  Within this model, the
oscillation at $\Gamma$ is undamped; however, external couplings would
be expected to cause dephasing.  The single mode contribution to the
dephasing of the $2 \Delta$ oscillation becomes significant when $p
(\equiv\alpha \frac{\Delta}{\omega}) >>1$ in which case $Q =
(\frac{\Delta}{\alpha \omega})^{1/2}$.  Using $\alpha = 1$ and the
adiabatic ratio $\Delta/\omega = 10$, yields a $Q$ for the qubit of
$Q\simeq 3.3$. As in fig. 1, however, the spectrum is discrete
consisting of several sidebands. In contrast, the ohmic spectrum,
$J(\omega) = \eta \omega$, for the same DC-SQUID parameters (and lead
resistance $R=100 \Omega$) has a dimensionless coupling of $\eta \sim
0.03$.  The ohmic contribution to the dephasing rate is $1/\tau \simeq
2 \pi \eta \left({\epsilon \over \Delta}\right)^2 {T \over \hbar}$,
which vanishes for both zero bias and zero temperature (the present
case).  For $\epsilon = \Delta \sim 1$ GHz, the ohmic $Q$ becomes
comparable for a temperature of $T \sim 10$ mK and diverges for $T=0$.
In contrast, the single mode contribution to dephasing remains
constant at $T=0$.

In conclusion, we have considered a model in which a qubit is strongly
coupled to a single environmental mode and described the resulting
dephasing and entanglement.  We have solved the resulting model by
appealing to a formal analogy with the Fluctuating Gap model, a model
for disordered fermions. We applied these results to an underdamped SQUID environment, however, this transition might be studied in a variety of existing qubit designs, the main attribute being a variable coupling between the qubit and an adiabatic ($\Delta/\omega \sim 10$) oscillator mode. 

We thank Jeffrey Clayhold, Boris Fine, Philip Stamp, and Oleg Starykh
for many useful discussions.

\end{document}